\documentclass[useAMS,usenatbib]{mn2e}
\usepackage{epsfig}
\usepackage{threeparttable}
\usepackage{amsmath}

\title[Low braking index of PSR J1734-3333]{Low braking index of PSR J1734-3333: an interaction between fall-back disk and magnetic field?}
\author[W. C. Chen, and X. D. Li]{ Wen-Cong Chen $^{1,3}$\thanks{E-mail:
chenwc@pku.edu.cn}, and Xiang-Dong Li $^{2,3}$,\\ $^1$ Department
of Physics, Shangqiu Normal University,
Shangqiu 476000, China;\\
 $^2$ Department of Astronomy, Nanjing University, Nanjing 210093, China;
 lixd@nju.edu.cn\\
 $^3$ Key Laboratory of Modern Astronomy and Astrophysics (Nanjing University), Ministry of Education, Nanjing 210093, China}
\begin{document}

\date{}

\pagerange{\pageref{firstpage}--\pageref{lastpage}} \pubyear{2011}

\maketitle

\label{firstpage}

\begin{abstract}
Recent timing observation reported that the radio pulsar PSR J1734
- 3333 with a rotating period $P=1.17~\rm s$ is slowing down with
a period derivative $\dot{P}=2.28\times 10^{-12}\rm s\,s^{-1}$.
Its derived braking index $n=0.9 \pm 0.2$ is the lowest value
among young radio pulsars with the measured braking indices. In
this Letter, we attempt to investigate the influence of the
braking torque caused by the interaction between the fall-back
disk and the strong magnetic field of the pulsar on the spin
evolution of PSR J1734 - 3333. Analytical result show that this
braking torque is obviously far more than that by magnetic dipole
radiation for pulsars with spin period of $> 0.1$ s, and play an
important role during the spin-down of the pulsars. Our simulated
results indicate that, for some typical neutron star parameters,
the braking index and the period derivative approximately in
agreement with the measured value of PSR J1734 - 3333 if the
material inflow rate in the fallback disk is $2 \times 10^{17} \rm
g\,s^{-1}$. In addition, our scenario can account for the measured
braking indices of four young pulsars. However, our predicted X-ray luminosity are 1 -2
order of magnitude higher than the observation. We proposed that this discrepancy
may originate from the instability of fall-back disk.
\end{abstract}


\section{Introduction}
Radio pulsars are believed to be rapid rotating, strongly
magnetized neutron stars, which originated from the core collapse
of massive stars during supernova explosions \citep{paci67}. By
magnetic dipole radiation or charged particle winds, pulsars lose
rotational kinetic energy, and their spin periods gradually
increase \citep{gold68}. In the diagram of pulse period ($P$)
versus period derivative ($\dot{P}$), normal radio pulsars locate
a concentrated region, which has periods of $\sim~1\rm s$ and
period derivatives of $10^{-16} - 10^{-14}~\rm s\,s^{-1}$
\citep{man04}. Recent timing observation reported that radio
pulsar PSR J1734 - 3333 with a rotating period $P=1.17~\rm s$ is
slowing down with a relatively high period derivative
$\dot{P}=2.28\times 10^{-12}\rm s\,s^{-1}$ \citep{espi11}. It is
exciting that this source has a lowest braking index
$n=\Omega\ddot{\Omega}/\dot{\Omega}^{2}=0.9\pm0.2$ (where
$\Omega=2\pi/P$) among young pulsars with the measured braking
indices.

The braking index of pulsars is determined by the slow-down
torque. If the magnetic fields of pulsars are constant, magnetic
dipole radiation predicted the braking index $n=3$. However, all
the measured braking indices for young radio pulsars are less than
3. \cite{blan88} suggested that the variation of magnetic moment
may cause $n<3$. The braking torque caused by the magnetic dipole
radiation and the unipolar generator may produce the braking
indices of 1 - 3 \citep{xu01}. \cite{taur01} argued that the
exponential models for magnetic field decay and alignment can
account for the observed data. Adopting a magnetic field evolution
model, \cite{zhang12} can interpret all the observed statistical
properties of the braking indices of the pulsars in \cite{hobb10}.
Furthermore, other braking torques may result in the braking
indices $n<3$. As a dominant spin-down torque, strong relativistic
winds give rise to $n=1$ \citep{mich69}. The propeller torque
caused by a fall-back disk of pulsars may be responsible for the
low braking index \citep{alpa01,meno01}. The tidal torque
originated from the gravitational interaction between the pulsar
and the fall-back disk can lead to the observed braking indices
\citep{chen06}. Recently, \cite{maga12} have modified canonical
model to explain the observed braking indices ranges, and then
predicted the possible values of the braking indices of several
other young pulsars.

The relatively small braking index of PSR J1734 - 3333 challenges
the present braking theories of pulsars. Assuming that the
magnetic field and the particle luminosity are both constants,
this source may be braking by a rotation-powered particle wind
\citep{espi11,tong13}. Adopting a modified formula for the
propeller torque, a self-similar fall-back disk can account for
the small braking index, $P, \dot{P}$ of PSR J1734 - 3333
\citep{liu14}. Using the same model that used to study the
evolution of anomalous X-ray pulsars and soft gamma-ray repeaters
\citep{erta09,alpa11}, a fall-back disk around PSR J1734 - 3333
can fit the observed period, the first and second period
derivatives, and the X-ray luminosity of this source
\citep{cali13}.

In this Letter, the braking torque caused by the interaction
between the strong magnetic field and the fall-back disk around
the pulsar is applied to investigate the evolution of PSR J1734 -
3333. In section 2, we describe the theoretical model of
interaction between the magnetic field and the fall-back disk. In
section 3, the calculated results for PSR J1734 - 3333 and other
six pulsars with known braking indices are presented. Finally, we
summarize the results with a brief discussion in section 4.
\section{Model}
A small amount of fall back material may form a debris disk (the
so-called fall-back disk) around the radio pulsars because all the
mass could not be fully ejected during the supernova explosion
\citep{mich88}. The brightest known AXP 4U 0142+61 was detected
the mid-infrared emission, which hints the existence of a
fall-back disk \citep{wang06}. Recently, the counterpart to AXP 1E
2259+586 at 4.5 $\rm \mu m$ can be interpreted by a X-ray-heated
fall-back disk \citep{kapl09}. PSR J1734-3333 is a young radio
pulsar, and may associated with the supernova remnant G354.8-0.8
\citep{man02}. It is possible that this source will evolve to the
magnetars region after about 30 kyr \citep{espi11}. Therefore, PSR
J1734-3333 may has a fall-back disk that has not yet been
detected.

During the evolution of pulsars, the braking torque originating
from the magnetic coupling between the magnetic field and the
fall-back disk should play an important role. The magnetic field
lines of the pulsar could penetrate the fall-back disk, and then
twist due to the differential rotation between the pulsar and the
disk \citep{ghos79a,ghos79b}. The resulting magnetic torque
($N_{\rm mag}$) would transfer angular momentum between the pulsar
and the disk, and spin down or up the pulsar. The magnetic torque
$N_{\rm mag}$ depends on the "fastness parameter"
$\omega=\Omega/\Omega_{\rm m}$, where $\Omega, \Omega_{\rm m}$ are
the angular velocity of the pulsar and the inner radius of the
fall-back disk, respectively. In this work, the inner radius of
the fall-back disk is taken to be the magnetosphere radius
\begin{equation}
R_{\rm m}=1.6\times 10^{8}\left(\frac{B}{10^{12}\rm
G}\right)^{4/7}\left(\frac{\dot{M}}{10^{18}\rm
g\,s^{-1}}\right)^{-2/7}\rm cm.
\end{equation}
For radio pulsar, accretion process can not occur, and the
magnetosphere radius is larger than the co-rotation radius $R_{\rm
co}=(GM/\Omega^{2})^{1/3}$. Therefore, $\Omega>\Omega_{\rm m}$,
and the "fastness parameter" $\omega>1$. Adopting some typical
assumptions, the magnetic torque can be written as \citep[see
also][]{dai06}
\begin{equation}
N_{\rm mag}=\frac{\dot{M}\sqrt{2GMR_{\rm
m}}}{3}(\frac{2}{3\omega}-1),
\end{equation}
where $G$ is the gravitational constant, $M$ the mass of the
pulsar, $\dot{M}$ the mass inflow rate in the fall-back disk.

We assume that PSR J1734-3333 is a radio pulsar with a normal
magnetic field. It total braking torque consists of magnetic
dipole radiation and the magnetic torque, i. e.
\begin{equation}
I\dot{\Omega}=N_{\rm dip}+N_{\rm mag},
\end{equation}
where $I$ is the momentum of inertia of the pulsar. The torque
caused by the magnetic dipole radiation
\begin{equation}
N_{\rm dip}=-\frac{2B^{2}R^{6}{\rm
sin}^{2}\theta}{3c^{3}}\Omega^{3},
\end{equation}
where $B, R$ are the surface magnetic field, and the radius of the
pulsar, respectively; $\theta$ is the inclination of the magnetic
axis with respect to the rotation axis of the pulsar.

\section{Results}

Adopting some typical parameters for the pulsar as follows,
$B=10^{12}~{\rm G}, \theta=\pi/2, M=1.4~{\rm M_{\odot}},
R=10^{6}~\rm cm$, we have $N_{\rm dip}=-2.47\times 10^{28}~\rm
g\,cm^{2}s~\Omega^{3}$. If the mass inflow rate in the fall-back
disk $\dot{M}=1\times10^{17}\rm g\,s^{-1}$, we can derived $R_{\rm
m}=3.09\times 10^{8}~\rm cm$, $\Omega_{\rm m}=2.5~\rm s^{-1}$, and
$N_{\rm mag}=-1.1\times 10^{34}(1-5.0/(3\Omega))~\rm
g\,cm^{2}s^{-2}$. In figure 1, we plot the braking torque of the
pulsars as a function of the spin period. It can be clearly seen
that, for nascent pulsars with a rapid spin, the braking torque by
magnetic dipole radiation is dominant. With the spin-down, the
magnetic torque gradually exceed the magnetic dipole radiation
torque at $P\approx 0.08 - 1.1$ ms, and then dominate the
spin-down of pulsars.

\subsection{PSR J1734-3333}

It is clear that, $|N_{\rm mag}|\gg |N_{\rm dip}|$ for PSR
J1734-3333, hence we can neglect the torque caused by the magnetic
dipole radiation. Assuming that the mass inflow rate in the
fall-back disk and the magnetic field are constant, the braking
index of PSR J1734-3333 is given by
\begin{equation}
n=\frac{\ddot{\Omega}\Omega}{\dot{\Omega}^{2}}=\frac{2\Omega_{\rm
m}}{3\Omega-2\Omega_{\rm m}}.
\end{equation}
The period derivative can be written as
\begin{equation}
\dot{P}=-\frac{N_{\rm mag}P^{2}}{2\pi I},
\end{equation}
and the second order period derivative is
\begin{equation}
\ddot{P}=\frac{2\dot{P}^{2}}{P}-\frac{\sqrt{2GMR_{\rm
m}}\dot{M}}{3 I}\frac{2\Omega_{\rm m}}{3\Omega^{2}}\dot{P}.
\end{equation}

\begin{table}
\begin{center}
\centering \caption{ Observed and derived parameters for PSR J1734
- 3333.\label{tbl-1}}
\begin{tabular}{cc}
\hline
Parameter &  Value \\
\hline
$P$(s)& 1.17\\
$\dot{P}(\rm s\,s^{-1})$&$2.28\times 10^{-12}$\\
$\ddot{P}(\rm s\,s^{-2})$& $5.0\pm0.8\times 10^{-24}$ \\
Characteristic age (kye) & 8.1\\
Surface magnetic field (G) & $5.2\times 10^{13}$\\
Braking index, n& $0.9\pm 0.2$\\

 \hline
\end{tabular}

\end{center}
\end{table}

Table 2 summarizes the calculated braking index, $\dot{P}$, and
$\ddot{P}$ for  PSR J1734 - 3333. As $\dot{M}=2\times10^{17}\rm
g\,s^{-1}$, the predicted braking index can fit the observation.
However, there exist $\sim10\%$ and 50\% errors between the
predicted values and the observations for $\dot{P}$, and
$\ddot{P}$, respectively.


\begin{table*}
\centering
\begin{minipage}{100mm}
\caption{Main results of the fall-back disk model for PSR J1734 -
3333.}
\begin{tabular}{@{}lllllll@{}}
  \hline\hline\noalign{\smallskip}
$\dot{M}(\rm g\,s^{-1})$&$R_{\rm m}(\rm cm)$&$\Omega_{\rm m}$($\rm s^{-1}$)&$\dot{P}$($\rm s\, s^{-1}$)&$\ddot{P}$($\rm s\, s^{-2}$)&$n$\\
  \hline\noalign{\smallskip}
$1\times10^{17}$ &$3.09\times 10^{8}$&2.5& $1.70\times10^{-12}$&$3.8 \times 10^{-24}$ & 0.45    &\\
$2\times10^{17}$ &$2.53\times 10^{8}$&3.4& $2.57\times10^{-12}$&$7.2 \times 10^{-24}$ & 0.73    &\\
$3\times10^{17}$ &$2.26\times 10^{8}$&4.0& $3.18\times10^{-12}$&$8.7 \times 10^{-24}$ & 0.99    &\\
\hline
$5\times10^{13}$ &$2.71\times 10^{9}$&0.097& $3.59\times10^{-15}$&$2.2 \times 10^{-29}$ & 0.57    &\\
\noalign{\smallskip}\hline
\end{tabular}
\end{minipage}
\end{table*}

\subsection{Other pulsars with known braking indices}

In order to test our model, we also calculated the braking indices
of other six pulsars in Table 3. Figure 1 shows that, for pulsars
with spin period $\la 0.1$ s the braking torque by magnetic dipole
radiation cannot be ignored. Therefore, we can derived the general
braking index from equation (3)
\begin{equation}
n=3-\frac{S(\frac{8\Omega_{\rm m
}}{3\Omega}-3)}{-K\Omega^{3}+S(\frac{2\Omega_{\rm
m}}{3\Omega}-1)},
\end{equation}
where $K=2B^{2}R^{6}{\rm sin}^{2}\theta/(3c^{3})$,
$S=\dot{M}\sqrt{2GMR_{\rm m}}/3$. Taking the typical parameters of
pulsars, $K=2.47\times 10^{28}~\rm g\,cm^{2}s$, and $S$ can be
derived by the mass inflow rate $\dot{M}$ in the fall-back disk.

Figure 2 plots the evolutionary paths in the braking indices vs.
spin periods of pulsars surrounded by a fall-back disk. Since
magnetic dipole radiation is the dominant spin-down mechanism, the
braking indices of pulsars with rapid spin ($P\la 0.03$ s when $\dot{M}\sim
10^{17}~\rm g\,s^{-1}$, and $P\la 0.3$ s when $\dot{M}\sim
10^{13}-10^{14}~\rm g\,s^{-1}$, ) are
very near 3. With the sharp increase of the magnetic torque, the
braking index firstly decrease to a minimum of 0.2 at $P\sim 0.2 -
0.6$ s  when $\dot{M}\sim
10^{17}~\rm g\,s^{-1}$, and then sharply increase in the last stage. To compare
with observations, we also show the locations of seven pulsars
with known braking indices by the solid squares in Figure 2. As
shown in this figure, our simulated results can roughly fit the
observed braking indices of four pulsars  when $\dot{M}\sim
10^{17}~\rm g\,s^{-1}$. For a lower mass inflow rate $\dot{M}\sim
10^{13}-10^{14}~\rm g\,s^{-1}$, our simulation can approximately account for
the braking indices of PSR J1734 -
3333 and other 3 pulsars. However, the magnetic torque $N_{\rm mag}\propto \dot{M}^{6/7}$,
so the inferred $\dot{P}$ and $\ddot{P}$ of PSR J1734 - 3333 are obviously
lower than the observation (see Table 2). Table 3 shows that the period derivative of other three
pulsars are also $\sim 10^{-12}~\rm s\,s^{-1}$. Therefore, it is difficult for the fallback disk
with a low mass inflow rate to explain the braking process of pulsars.

\begin{figure}
 \includegraphics[width=0.53\textwidth]{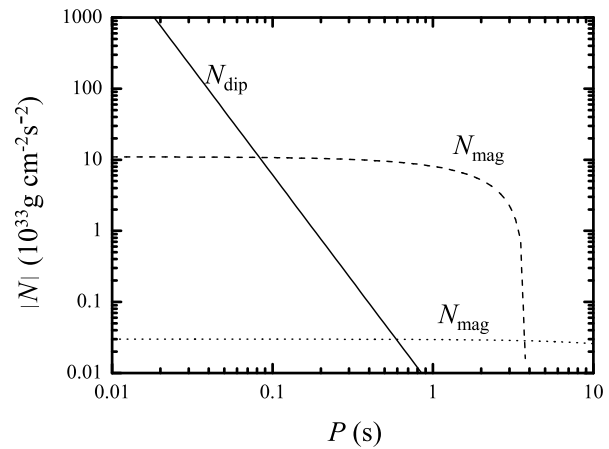}
  \caption{ Braking torques of pulsars as a function of the spin
  period. The solid, dashed, and dotted curves denote the braking torque caused by
  magnetic dipole radiation and the magnetic torques when
the mass inflow rate in the fall-back disk $\dot{M}=
1\times10^{17}, 1\times10^{14}~\rm g\,s^{-1}$, respectively.}
 \label{fig:fits}
\end{figure}

\begin{table}
\begin{center}
\centering \caption{ Spin periods and braking indices for seven
pulsars.\label{tbl-1}}
\begin{tabular}{ccccc}
\hline
Pulsar &  $P$  & $\dot{P}$&$n$  & Reference\\
       &  (s)        &     ($10^{-12}\rm s\,s^{-1}$)           &         &       \\
\hline
Crab& 0.0334  &  0.42  &2.509 & 1\\
PSR B0540-69 & 0.0505 & 0.48 & 2.14& 2,3 \\
Vela & 0.0893 &0.13 &$1.4\pm 0.2$ & 4\\
PSR B1509-58 &  0.151 & 1.5 & 2.837 &2 \\
PSR J1846-0258 & 0.327& 7.1  &2.16 & 5\\
PSR J1119-6127 & 0.408&  4.0  & 2.684 & 6\\
PSR J1734-3333 &1.17 & 2.3 &$0.9\pm 0.2$ & 7\\
 \hline
\end{tabular}
  \begin{tablenotes}
Reference. (1) \cite{lyne93}; (2) \cite{livi07}; (3)
\cite{boyd95}; (4) \cite{lyne96}; (5) \cite{livi11}; (6)
\cite{walt11}; (7) \cite{espi11}.
   \end{tablenotes}
\end{center}
\end{table}

\begin{figure}
 \includegraphics[width=0.53\textwidth]{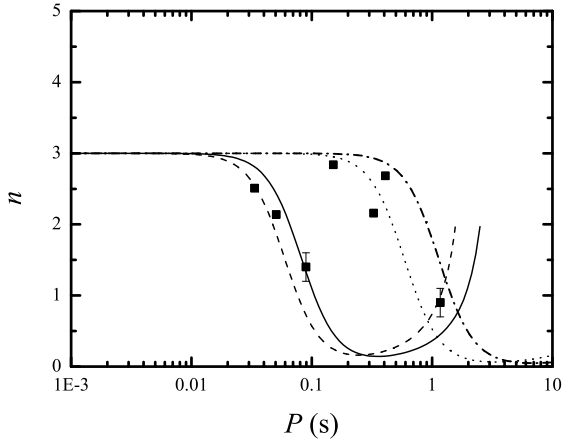}
  \caption{Braking index of pulsars as a function of the spin
  period. The solid, dashed, dotted, and dashed-dotted curves represent the braking index when
the mass inflow rate in the fall-back disk
$\dot{M}=1\times10^{17}$, $3\times10^{17}$, $10^{14}$, and $10^{13}~\rm g\,s^{-1}$
respectively. The solid squares represent seven pulsars with known
braking indices.}
 \label{fig:fits}
\end{figure}

\section{Discussion and summary}
In this Letter, we proposed an alternative scenario to interpret
the low braking index of PSR J1734-3333. In our opinion, this
pulsar should has a normal magnetic field. However, it may be
surrounded by a fall-back disk, and the magnetic torque
originating from the magnetic coupling between the magnetic field
and the fall-back disk can efficiently brake the pulsar. Our
calculations show that, for PSR J1734-3333 the magnetic torque is
obviously far more than the braking torque by magnetic dipole
radiation. When the material inflow rate in the fallback disk is
$2 \times 10^{17} \rm g\,s^{-1}$, our calculated braking index and
period derivative approximately in agreement with the measured
value.

However, our scenario present a relatively high X-ray luminosity.
The X-ray luminosity of the fall-back disk can be calculated by
$L_{\rm X}\sim \frac{GM\dot{M}}{2R_{\rm m}}
\approx 7.4\times 10^{34}\rm erg\,s^{-1}$, which is obviously higher the
the inferred X-ray luminosity (0.5 - 10 keV, $0.1 - 3.4\times 10^{33}\rm erg\,s^{-1}$)
of PSR J1734-3333 \citep{olau10}. This discrepancy
may originate from the instability of fall-back disk. The inflow plasma
with not enough kinetic energy can not escape from the system, and build-up in the disk
\citep{dang10,dang11,dang12}, which experienced short
outbursts separated by long quiescent intervals (similar to transient X-ray source).
In a recurrent period, the average angular velocity derivative
$\dot{\Omega}=\dot{\Omega}_{h}d+\dot{\Omega}_{l}(1-d)$, here $d$ is the duty cycle,
and $\dot{\Omega}_{h}, \dot{\Omega}_{l}$
can be derived by equation (3) when the pulsar is in high state and low state, respectively.
The constant mass inflow rate adopted in section 3 is the average value in a recurrent period. 
Therefore, the spin evolution of the pulsar is mainly influenced by the mass inflow rate in high state.
The mass inflow rate in low state with long interval is very low, and the neutron star can be
observed as a radio pulsar. If so, PSR J1734 - 3333 may be a intermittent pulsar.

Our model is also applied in other six pulsars with known braking
indices. Including PSR J1734-3333, our predicted braking indices
can fit the observed values of four pulsars. However, our
calculations show that pulsars with $P\sim 0.2 - 0.6$ s and normal
magnetic field ($\sim 10^{12}$ G) may have ultra-low braking
indices of $\sim 0.2$. We expect further detailed radio
observations for pulsars to confirm or negate our idea in the
future. It is noticed that our scenario cannot fit the observed
braking indices of other three pulsars. We suspect they may have
relatively strong magnetic fields. Some
pulsars might be born with strong magnetic fields or enhance their
magnetic fields by strong glitches \citep{lin04}.

\section*{Acknowledgments}
We are grateful to the anonymous referee for helpful comments.
This work was partly supported by the National Science Foundation
of China (under grant number 11173018, and 11573016), and Innovation Scientists
and Technicians Troop Construction Projects of Henan Province,
China.

\bsp

\label{lastpage}


\begin{thebibliography}{99}
\bibitem[\protect\citeauthoryear{Alpar, Ankay \& Yazgan}{1988}]{alpa01} Alpar M. A., Ankay A., Yazgan E., 2001, ApJ, 557, L61
\bibitem[\protect\citeauthoryear{Alpar, Ertan \& \c{C}ali\c{s}kan}{2011}]{alpa11} Alpar M. A., Ertan \"{U}., \c{C}ali\c{s}kan \c{S}., 2011, ApJ, 732, L4
\bibitem[\protect\citeauthoryear{Blandford \& Romani}{1988}]{blan88} Blandford R. D., Romani R. W., 1988, MNRAS, 234, 57
\bibitem[\protect\citeauthoryear{\c{C}ali\c{s}kan et al.}{2013}]{cali13} \c{C}ali\c{s}kan \c{S}., Ertan \"{U}., Alpar M. A., Tr\"{u}mper J.
E., Kylafis N. D., 2013, MNRAS, 431, 1136
\bibitem[\protect\citeauthoryear{Boyd et al.}{2006}]{boyd95} Boyd P. T., van Citters G. W., Dolan J. F., et al. 1995, ApJ, 448, 365
\bibitem[\protect\citeauthoryear{Chen \& Li}{2006}]{chen06} Chen W. C., Li X. D., 2006, A\&A, 450, L1
\bibitem[\protect\citeauthoryear{Dai \& Li}{2006}]{dai06} Dai H.-L., Li X.-D., 2006, A\&A, 451, 581
\bibitem[\protect\citeauthoryear{D'Angelo \& Spruit}{2010}]{dang10}D'Angelo C. R., Spruit H. C., 2010, MNRAS, 406, 1208
\bibitem[\protect\citeauthoryear{D'Angelo \& Spruit}{2011}]{dang11}D'Angelo C. R., Spruit H. C., 2011, MNRAS, 416, 893
\bibitem[\protect\citeauthoryear{D'Angelo \& Spruit}{2012}]{dang12}D'Angelo C. R., Spruit H. C., 2012, MNRAS, 420, 416
\bibitem[\protect\citeauthoryear{Ertan et al.}{2009}]{erta09} Ertan \"{U}., Ek\c{s}i K. Y., Erkut M. H., Alpar M. A., 2009, ApJ, 702, 1309
\bibitem[\protect\citeauthoryear{Espinoza et al.}{2011}]{espi11} Espinoza C. M., Lyne A. G., Kramer M., et al., 2011, ApJL, 741, L13
\bibitem[\protect\citeauthoryear{Gold}{1968}]{gold68} Gold T., 1968, Nat, 218, 731
\bibitem[\protect\citeauthoryear{Ghosh \& Lamb}{1979a}]{ghos79a} Ghosh P., Lamb F. K., 1979a, ApJ, 232, 259
\bibitem[\protect\citeauthoryear{Ghosh \& Lamb}{1979b}]{ghos79b} Ghosh P., Lamb F. K., 1979b, ApJ, 234, 296
\bibitem[\protect\citeauthoryear{Hobbs, Lyne \& Kramer}{2010}]{hobb10} Hobbs G., Lyne A. G., Kramer M. 2010, MNRAS, 402, 1027
\bibitem[\protect\citeauthoryear{Kaplan et al.}{2009}]{kapl09} Kaplan D. L., Chakrabarty D., Wang Z., Wachter, S., 2009, ApJ, 700, 149
\bibitem[\protect\citeauthoryear{Lin \& Zhang}{2004}]{lin04} Lin J. R., Zhang S. N. 2004, ApJ, 615, L133
\bibitem[\protect\citeauthoryear{Liu et al.}{2014}]{liu14} Liu X. -W., Xu R.-X., Qiao G. -J., Han J. -L., Tong H., 2014, RAA, 14, 85
\bibitem[\protect\citeauthoryear{Livingstone et al.}{2007}]{livi07} Livingstone M. A., Kaspi V. M., Gavriil F. P., et al. 2007,
Ap\&SS, 308, 317
\bibitem[\protect\citeauthoryear{Livingstone et al.}{2011}]{livi11} Livingstone M. A., Ng C.-Y., Kaspi V. M.,
Gavriil F. P., Gotthelf E. V. 2011, ApJ, 730, 66
\bibitem[\protect\citeauthoryear{Lyne, Pritchard \& Graham Smith}{1993}]{lyne93} Lyne A. G., Pritchard R. S.,  Graham Smith F. 1993, MNRAS, 265, 1003
\bibitem[\protect\citeauthoryear{Lyne et al.}{1996}]{lyne96} Lyne A. G., Pritchard R. S., Graham Smith F., Camilo, F. 1996,
Nat, 381, 497
\bibitem[\protect\citeauthoryear{Magalhaes, Miranda \& Frajuca}{2012}]{maga12} Magalhaes, N. S., Miranda, T. A., Frajuca, C. 2012, ApJ, 755, 54
\bibitem[\protect\citeauthoryear{Manchester et al.}{2002}]{man02} Manchester R. N., Bell J. F., Camilo F., et al., 2002, in ASP
Conf. Ser. 271, Neutron Stars in Supernova Remnants, ed. P. O.
Slane \& B. M. Gaensler (San Francisco, CA: ASP), 31
\bibitem[\protect\citeauthoryear{Manchester}{2004}]{man04} Manchester R. N., 2004, Sci, 304, 542
\bibitem[\protect\citeauthoryear{Menou, Perna \& Hernquist }{2001}]{meno01} Menou K., Perna R., Hernquist L., 2001, ApJ, 554, L63
\bibitem[\protect\citeauthoryear{Michel}{1969}]{mich69} Michel F. C., 1969, ApJ, 158, 727
\bibitem[\protect\citeauthoryear{Michel}{1988}]{mich88} Michel F. C., 1988, Nature, 333, 644
\bibitem[\protect\citeauthoryear{Olausen et al.}{2010}]{olau10} Olausen S. A., Kaspi V. M., Lyne A. G., Kramer M. 2010, ApJ, 725, 985
\bibitem[\protect\citeauthoryear{Pacini}{1967}]{paci67} Pacini F., 1967, Nat, 216, 567
\bibitem[\protect\citeauthoryear{Papitto et al.}{2015}]{papi15} Papitto A., de Martino D., Belloni T. M., et al. 2015, MNRAS, 449, L26
\bibitem[\protect\citeauthoryear{Tauris \& Konar}{2001}]{taur01} Tauris T. M., Konar S., 2001, A\&A, 376, 543
\bibitem[\protect\citeauthoryear{Tong et al.}{2013}]{tong13} Tong H., Xu R. X., Song L. M., Qiao G. J., 2013, ApJ, 768, 144
\bibitem[\protect\citeauthoryear{Waltevrede, Johnston \& Espinoza }{2011}]{walt11} Waltevrede P., Johnston S., Espinoza C. M. 2011, MNRAS, 411, 1917
\bibitem[\protect\citeauthoryear{Wang et al.}{2006}]{wang06} Wang Z., Chakrabarty D., Kaplan D. L., 2006, Nat, 440, 772
\bibitem[\protect\citeauthoryear{Xu \& Qiao}{2001}]{xu01} Xu R. X., Qiao G. J., 2001, ApJ, 561, L85
\bibitem[\protect\citeauthoryear{Zhang}{2005}]{zhang05} Zhang S. -N., 2005, ApJ, 618, L79
\bibitem[\protect\citeauthoryear{Zhang \& Xie}{2012}]{zhang12} Zhang S. -N., Xie Y., 2012, ApJ, 761, 102





\end{thebibliography}
\end{document}